\documentclass[aps,pra,floatfix,twocolumn]{revtex4}
\usepackage{graphicx,amsmath}

\begin{document}

\title{Nonexistence of Entanglement Sudden Death in High NOON States
\vspace{-2mm}}

\author{Asma Al-Qasimi, Daniel F. V. James}
\affiliation{Department of Physics, University of Toronto, Toronto ON M5S 1A7, Canada.
\vspace{-2mm}}

\begin{abstract} \noindent
We study the dynamics of entanglement in continuous variable quantum systems (CVQS). Specifically, we study the phenomena of Entanglement Sudden Death (ESD) in general two-mode-N-photon states undergoing pure dephasing. We show that for these states, ESD \emph{never} occurs. These states are generalizations of the so-called High NOON states, shown to decrease the Rayleigh limit of $\lambda$ to $\lambda/N$, which promises great improvement in resolution of interference patterns if states with large $N$ are physically realized \cite{Dowling}. However, we show that in dephasing NOON states, the time to reach $V_{crit}$, critical visibility, scales inversely with $N^{2}$. On the practical level, this shows that as $N$ increases, the visibility degrades much faster, which is likely to be a considerable drawback for any practical application of these states.
\end{abstract}

\maketitle

Enanglement is a quantum property that, for a long time, has fascinated those studying the fundamentals of quantum mechanics, and, more recently those interested in its powerful applications such as Quantum Information science. Many argue that not only is it \emph{a} quantum property, but rather the \emph{only} one. The question of when entanglement disappears is an interesting fundamental question to consider.

Entanglement Sudden Death is a term coined by Yu and Eberly \cite{YuEberly} to decribe loss of entanglement in a finite time. The work done so far, which mostly concerns two-qubit systems, has shown, in one way, how fragile entanglement is in realistic systems. Several papers have shown that ESD always occurs in some very general two qubit systems. Examples include X-states, states with nonzero parameters (in general) on the diagonal and antidiagonal of the density matrix of the system. In \cite{YEDephasing}, it is shown that for dephasing in X-states, there is always ESD as long as none of the parameters of the density matrix are zero, and in \cite{AsmaDaniel}, it is shown that for these states at finite temperature, and with depopulation going on, ESD also always occurs.
With results like these, one is tempted to make the guess that ESD is actually a universal phenomena. So far an attempt has been made to prove this in \cite{YEProof}.

With all this work showing how prevalent ESD is in qubit systems, it is interesting to ask how common it is in other quantum systems. For example, is ESD as common in Continuous Variable Quantum Systems (CVQS) as it is in qubits? Recently, ESD has been shown to occur in a system of two free harmonic oscillators interacting with a Markovian bath \cite{ESDCVQS}. In addition, two initially Gaussian states, states with Gaussian Wigner functions, coupled to the same (ohmic) environment have been studied in \cite{Paz}, where the existence of three phases where demonstrated: ESD, ESD with revival, and No ESD. Entanglement Sudden Death in CVQS systems has also been studied taking into consideration relativistic effects \cite{relCVQS}.

In the work we present here, we prove the general result that ESD never occurs in two-mode-N-photon states undergoing \emph{dephasing}. NOON states, which have been shown \cite{Dowling} to beat the Rayleigh limit in interferometry, falls under this general class of states. The resolution of interference patterns improves when the separation between the wave amplitudes falls down to $\lambda/N$ compared to the minimum of $\lambda$ forced by the Rayleigh limit. The power of these NOON states lies in their entanglment. It is, therefore, important to study the decay of entanglement in these systems. The approach we describe here is studying ESD in such systems, trying to get a feel for the fragility of entanglement. Finally, we touch on the important question: Does the existence of some entanglement for a very long time have any practical implication on the usefulness of NOON states?

The system we consider is that of two harmonic oscillators with N photons shared between them. The density matrix, most generally, describing such a system is given as follows:

\begin{eqnarray}
\hat{\rho}(t) &=& \sum^{N}_{k=0}\rho_{kk}(t)\left|N-k,k\right\rangle \left\langle N-k,k\right| \nonumber \\
&+& \sum^{N}_{\substack{k,m=0, \\ k\neq m}}\rho_{km}(t)\left|N-k,k\right\rangle\left\langle N-m,m\right|.
\label{a}
\end{eqnarray}

\noindent We deal only with dephasing, since applications of such states tend to be post-selective on photon number: processes in which N changes are filtered away. If the system undergoes pure dephasing, due to random fluctuation of the mode frequency, one expects that the off-diagonal terms; i.e., $\rho_{km}(t)$, where $k\neq m$, in (\ref{a}) will acquire decay terms. On the other hand, the population, which is represented by the diagonal elements will remain intact; i.e., the photon number will be preserved. It can be easily checked that the Master equation describing the dynamics of such a system, in which there is no correlation between the two fields interacting with the two harmonic oscillators, is given by:

\begin{eqnarray}
\frac{\partial \hat{\rho}(t) }{\partial t}&=& 2\left(\Gamma_1 \left\{\left[\hat{n_1},\hat{\rho}(t)\hat{n_1}\right]+\left[\hat{n_1}\hat{\rho}(t),\hat{n_1}\right]\right\}\right. \nonumber \\
&&\left.\Gamma_2 \left\{\left[\hat{n_2},\hat{\rho}(t)\hat{n_2}\right]+\left[\hat{n_2}\hat{\rho}(t),\hat{n_2}\right]\right\}\right),
\label{a1}
\end{eqnarray}

\noindent where $\Gamma_i$ and $\hat{n_i}$ are the decay rate and the number state operator of the $i$th harmonic oscillator. By assumption there is no depopulation going on. By (\ref{a1}), the evolution of (\ref{a}) with respect to time in the case of pure dephasing is:

\begin{eqnarray}
&&\hat{\rho}(t) = \sum^{N}_{k=0}\rho_{kk}(0)\left|N-k,k\right\rangle \left\langle N-k,k\right| \nonumber \\
&+&\sum^{N}_{\substack{k,m=0, \\ k\neq m}}\rho_{km}(0)e^{-\frac{1}{2}\left(k-m\right)^{2}\left(\Gamma_{1}+\Gamma_{2}\right)t}\times \nonumber \\
&& \times \left|N-k,k\right\rangle\left\langle N-m,m\right|.
\label{a2}
\end{eqnarray}

As with other decoherence mechanisms, one side effect of dephasing is the decay of entanglement. To find out whether this decay results in ESD or not, we need to use a reliable measure for entanglement. In two-qubit systems, we have good measures such as Wootter's concurrence \cite{Wootters}, which can tell with certainty whether a system is entangled or separable. On the other hand, in the general two-CVQS, the best any measure can do is provide a necessary but not sufficient condition for separability \cite{Peres}. In the case of CVQS, the sufficiency has only been proven for Gaussian states \cite{Zoller}, and NOON states, the generalization of which we discuss here, do not belong to this class. Nevertheless, this weakness in the criteria does not have to disadvantage the study of entanglement, as we will show in our case. 

Here is a brief description for Peres's criterion for entanglement \cite{Peres}. The density matrix of a bipartite system may be written as:

\begin{equation}
\hat{\rho}(t) = \sum_{i}c_{i}\rho^{'}_{i}\otimes \rho^{''}_{i},
\label{b2}
\end{equation}

\noindent Taking the partial transpose over one of its subsystems one obtains:

\begin{equation}
\hat{\sigma}(t) = \sum_{i}c_{i}(\rho^{'}_{i})^{T}\otimes \rho^{''}_{i}.
\label{b}
\end{equation}

\noindent If $\sigma$ has at least one negative eigenvalue, then we know with certainty that the system is entangled. However, if none of the eigenvalues are negative, then the system could be entangled or separable. The consequence of this weakness to our study of ESD is that the existence of ESD cannot be proven with certainty, while its nonexistence \emph{can} be proven with certainty.

Using this criteria, we obtain the following for the partial transpose of our state in (\ref{a}):

\begin{eqnarray}
\hat{\sigma}(t) &=& \sum^{N}_{k=0}\rho_{kk}(t)\left|N-k,k\right\rangle \left\langle N-k,k\right| \nonumber \\
&+& \sum^{N}_{\substack{k,m=0, \\ k\neq m}}\rho_{km}(t)\left|N-k,m\right\rangle\left\langle N-m,k\right|.
\label{c}
\end{eqnarray}

Notice that since $k$ and $m$ in the second terms of the right hand side of (\ref{c}) are not equal, the total number of photons in the kets and bras are never equal to N. In other words, $N-k+m\neq N$ and $N-m+k\neq N$. Mathematically, this implies that each pair of the matrix elements $\left|N-k,m\right\rangle\left\langle N-m,k\right|$ and $\left|N-m,k\right\rangle\left\langle N-k,m\right|$ fall in a different subspace than each other as well as the space of the diagonal elements.

This breaks the problem of finding the eigenvalues of $\sigma$ into finding the eigenvalues of $\frac{N(N+1)}{2}$ matrices; the remaining eigenvalues are just the diagonal elements of $\sigma$. Each of these matrices has the following form:

\begin{eqnarray}
\left|\rho_{km}\right|&&\left\{e^{i\theta}\left|N-k,m\right\rangle \left\langle N-m,k\right|\right. \nonumber \\
&&\left.+ e^{-i\theta}\left|N-m,k\right\rangle\left\langle N-k,m\right|\right\},
\label{f}
\end{eqnarray}

\noindent where $\theta$ is the phase of the matrix element $\rho_{km}$, and the vertical bars $|...|$ represent the norm of the quantity they enclose. It can be easily shown that the following state:

\begin{equation}
\left|\psi\right\rangle = \frac{1}{\sqrt{2}}\left\{e^{i\theta}\left|N-k,m\right\rangle-\left|N-m,k\right\rangle\right\}
\label{g}
\end{equation}

\noindent is an eigenvector of (\ref{f}) with eigenvalue $-|\rho_{km}|$.

This means that (\ref{c}), has at least one negative eigenvalue as long as one of the $\rho_{km}$, and consequently $\rho_{mk}$, is nonzero; i.e., this is true as long as there is some coherence in the system. If there is not any other decoherence mechanism, such as depopulation, going on as well, this is always true; there will always be entanglement in the system for any finite time. In other words, \emph{for a general two-mode-N-photon state undergoing pure dephasing, there is no sudden death of entanglement}. This is the main result of this paper.

We demonstrate our results using the first realized NOON state \cite{Steinberg}; i.e., for a 2-mode-3-photon state, specifically given by:

\begin{equation}
\left|\psi\right\rangle = \frac{1}{\sqrt{2}}\left\{\left|N0\right\rangle+\left|0N\right\rangle\right\},
\label{first}
\end{equation}

\noindent where $N=3$, but more generally by:

\begin{equation}
\left|\psi\right\rangle = a\left|30\right\rangle+b\left|21\right\rangle+c\left|12\right\rangle+d\left|03\right\rangle,
\label{second}
\end{equation}

\noindent where $\left|a\right|^{2}+\left|b\right|^{2}+\left|c\right|^{2}+\left|d\right|^{2}$. Applying the arguments above, we find that for the partial transpose of the density matrix of this system, the negative eigenvalues are: $-|a||b|e^{-\frac{1}{2}\left(\Gamma_{1}+\Gamma_{2}\right)t}$,
$-|a||c|e^{-2 \left(\Gamma_{1}+\Gamma_{2}\right)t}$,
$-|a||d|e^{-\frac{9}{2}\left(\Gamma_{1}+\Gamma_{2}\right)t}$,
$-|b||c|e^{-\frac{1}{2}\left(\Gamma_{1}+\Gamma_{2}\right)t}$, 
$-|b||d|e^{-2 \left(\Gamma_{1}+\Gamma_{2}\right)t}$, and
$-|c||d|e^{-\frac{1}{2}\left(\Gamma_{1}+\Gamma_{2}\right)t}$. Each of them involve a decay term due to dephasing. However, they only become zero after an infinite amount of time that renders the negative exponential zero. Therefore, for any finite time, there is always entanglement in the system, so ESD does not occur.

\begin{figure}[!b]
\vspace{-3mm}
\includegraphics[angle=270,width=0.9\columnwidth]{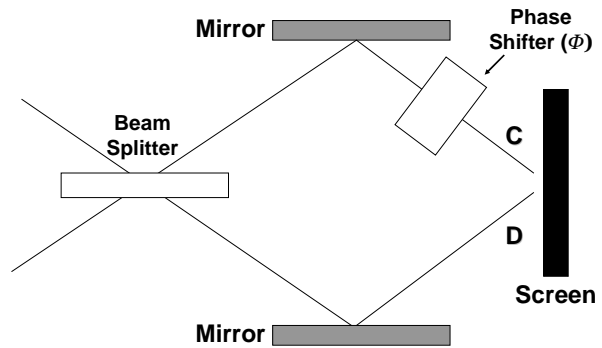}
\vspace{-2mm}
\caption{\textbf{Interference Pattern Formation} (\emph{Adapted from Figure 1 of \cite{Dowling}}) Two photon beams pass through a beam splitter and get reflected off the upper and lower mirrors to form an interference pattern on the screen. The upper beam passes through a phase shifter before reaching the screen. The phase aquired depends on the number of photons $N$ that pass through the upper path, and it equals $e^{iN\phi}$.}
\label{fig:figure1}
\end{figure}

\indent Finally, we consider how practical this long-lived entanglement is in NOON states undergoing dephasing. In Fig.~\ref{fig:figure1}, a standard setup for interfering two beams to produce interference fringes is described. The presence of a phase shifter (PS) in the upper path induces photons travelling there to acquire a phase shift $e^{i\phi}$. When the ``N-photon-NOON-state" is created inside this interferometer, the phase is acuumilated $N$ times, and the state becomes:

\begin{equation}
\left|\psi\right\rangle = \frac{1}{\sqrt{2}}\left\{\left|N0\right\rangle+e^{iN\phi}\left|0N\right\rangle\right\},
\label{DowlingNOON}
\end{equation}

\noindent With dephasing occuring, the density matrix of the state is as follows:

\begin{eqnarray}
\hat{\rho(t)}&&=\frac{1}{2}\left( \left|N0\right\rangle\left\langle N0\right|+e^{-N^{2}\Gamma t}e^{iN\phi}\left|0N\right\rangle\left\langle N0\right|\right. \nonumber \\
&&\left.+e^{-N^{2}\Gamma t}e^{-iN\phi}\left|N0\right\rangle\left\langle 0N\right|+\left|0N\right\rangle\left\langle 0N\right|\right),
\label{DephasedDowling}
\end{eqnarray}

\noindent where we assume $\Gamma_{1}=\Gamma_{2}=\Gamma$.

The expectaion value of the exposure dosage, $\langle\hat{\delta}\rangle$, displays fringes of visibility V. The exposure dosage operator $\hat{\delta}$ is described in terms of creation and annhilation operators acting on the two output (C and D paths) operators in Fig.~\ref{fig:figure1}. With simple algebra, it can be shown, in the case of dephasing of NOON states, that the expectation value of $\hat{\delta}$ is $\langle\hat{\delta}\rangle=1+e^{-N^{2}\Gamma t}cos(N\phi)$. From which the visibility is found to be:
\begin{equation}
V=\frac{\langle \hat{\delta} \rangle_{max}-\langle\hat{\delta}\rangle_{min}}{\langle\hat{\delta} \rangle_{max}+\langle \hat{\delta} \rangle_{min}}= e^{-N^{2}\Gamma t}
\label{visibility}
\end{equation}

\noindent When the visibility becomes vanishingly small, the fringes (and hence the measured phase) becomes impossible to measure. For the given decay rate $\Gamma$, the time it takes to reach $V_{crit}$, critical visibility, is given by
\begin{equation}
t_{crit}=\frac{1}{\Gamma N^{2}}ln\left(\frac{1}{V_{crit}}\right).
\label{time}
\end{equation}

\noindent Notice that in (\ref{time}), the expression for time depends inversely on $N^{2}$. This implies that the larger $N$ is, the faster it takes for visibilty to fall down to $V_{crit}$ and become worse. This is completely the opposite of what was earlier hoped to be acheived in improving resolution of fringes by creating High NOON states; i.e., states with large $N$.
 
\indent We showed that although the criteria for seperability has a weakness that can render some studies of entanglement uncertain, in our case and by using this criteria, we proved with \emph{certainty} that ESD does not occur in our system. We also demonstrate our result using the so called NOON states to show that there is no sudden death at NOON. Although this criteria allows us to prove that, even after a long time, there is \emph{some} enanglement left in the system, it does not give us a way to determine \emph{how much} is left. Therefore, to answer the question about the usefulness of the dephased NOON states in interferomentry, we study the time it takes to reach critical visibility. In doing so, we reveal that the presence of some entanglement does not have much practical implications. In fact, we show that for this realistic decohering system, increasing $N$ does not improve resolution, but rather allows it to worsen at a faster rate, which is proportionl to $N^{2}$.
This work was supported by NSERC.

\end{document}